\documentclass[a4paper,11pt]{article}
\usepackage{pos}
\usepackage{multicol}
\usepackage{amssymb,amsmath}
\title{The torelon spectrum and the world-sheet axion}

\newcommand{\be}{\begin{equation}}
\newcommand{\ee}{\end{equation}}
\newcommand{\bea}{\begin{eqnarray}}
\newcommand{\eea}{\end{eqnarray}}

\author*[a,b]{Andreas Athenodorou}
\author[c,d]{Michael Teper}

\affiliation[a]{Computation-based Science and Technology Research Center, The Cyprus Institute, Cyprus}
\affiliation[b]{Dipartimento di Fisica, Universit\'a di Pisa and INFN, Sezione di Pisa, Largo Pontecorvo 3, 56127 Pisa, Italy}
\affiliation[c]{Rudolf Peierls Centre for Theoretical Physics, University of Oxford, Parks Road, Oxford OX1 3PU, UK}
\affiliation[d]{All Souls College, University of Oxford, High Street, Oxford OX1 4AL, UK}

\emailAdd{a.athenodorou@cyi.ac.cy}
\emailAdd{mike.teper@physics.ox.ac.uk}

\abstract{We present a major update on the spectrum of the closed flux-tube (torelon) in $D=3+1$ $SU(N)$ gauge theories. Namely, we calculate the excitation spectrum of a confining flux-tube which winds around a spatial torus as a function of its length $l$, for short as well as long tubes. We do so for $N=3,5,6$ and two different values of the lattice spacing. Our states are characterised by the quantum numbers of spin $J$, transverse parity $P_{\perp}$, longitudinal parity $P_{\parallel}$ as well as by the longitudinal momentum $p_{\parallel}$. Our extended basis of operators used in combination with the generalized eigenvalue method enables us to extract masses for all irreducible representations characterised by $\{ |J|,P_{\perp},P_{\parallel} \}$.
We confirm that most of the low-lying states are well described by the spectrum of the Goddard–Goldstone–Rebbi–Thorn string. In addition we provide strong evidence, that in addition to string like states, massive modes exist on the world-sheet. More precisely the ground state with quantum numbers ${|J|}^{P_{\perp}, P_{\parallel}}=0^{--}$ exhibits a behaviour which is in agreement with the interpretation of being an axion on the world-sheet of the flux-tube. This state arises from a topological interaction term included in the effective world-sheet action. In addition we observe that the second excited state with ${|J|}^{P_{\perp}, P_{\parallel}}=0^{++}$ behaves as a massive mode with mass twice that of the axion.}

\FullConference{%
 The 38th International Symposium on Lattice Field Theory, LATTICE2021
  26th-30th July, 2021
  Zoom/Gather@Massachusetts Institute of Technology
}

\begin{document}
\maketitle

\section{Introduction}
\label{sec:introduction}
\vspace{-0.25cm}
In QCD quarks are confined in bound states by forming flux-tubes of chromo-magnetic and chromo-electric flux. Such a flux is connecting a quark with an anti-quark to form a bound state called an open-flux tube, which in nature manifests as a meson state. Long flux-tubes behave pretty much like thin strings: If you pull the string apart, at some point it breaks. However, to observe such a phenomenon in a lattice QCD calculation it requires the existence of dynamical quarks. Since we work on a pure gauge set up such effects do not appear. By placing the flux-tube in a particular position in space we expect $D-2$ massless modes to propagate along the string arising from the spontaneously broken translation invariance in the $D-2$ directions transverse to the flux. We, therefore, expect that there should be a low energy effective string theory describing such oscillating modes. Although, a flux-tube can be considered effectively as a string, it also has an intrinsic width. This suggests that massive states such as breather modes related to the intrinsic structure of the tube may exist in the spectrum. To investigate, whether, such states exist one needs to extract the aforementioned spectrum, compare it with an effective string theory model and identify states which exhibit significant deviations from the theoretical description. A naive expectation would be that a massive mode has the characteristics of a resonance; a constant term of the order of the mass gap ($\sim m_G$) coupled to a string state.

A decade ago we demonstrated~\cite{Athenodorou:2011rx} that the closed flux-tube spectrum in $D=2+1$ $SU(N)$ gauge theories can be well-approximated by the Nambu-Goto free string in flat space-time, from short to long flux-tubes, with no signs of any massive excitations. In addition we demonstrated~\cite{Athenodorou:2010cs} that the spectrum of the closed flux-tube in $D=3+1$ consists mostly of string-like states, however, in contrast to $D=2+1$ a number of states with quantum number $0^-$ appeared to encode the characteristics of a massive excitation. In 2013, Dubovsky et al,~\cite{Dubovsky:2013gi} demonstrated that this state arises naturally if one includes a Polyakov topological piece in the string theoretical action. Nevertheless, our previous results were poor - spectrum has been extracted for a few string lengths, and for low statistics. As a result, we could only extract states for the very low lying spectrum and for a restricted sector of the irreducible representations expanded by the quantum numbers (QNs) $\{ |J|,P_{\perp},P_{\parallel} \}$.

In this work we present a major improvement of our previous investigation on $D=3+1$. This has been achieved by increasing the basis of operators used to materialise the General Eigen-value Problem method \cite{Luscher:1990ck}, by extracting the spectrum of the flux-tube for three values of color $N$, namely $N=3,5,6$, as well as by probing through a large set of flux-tube lengths. In more detail, for each different set of quantum numbers $\{ |J|,P_{\perp},P_{\parallel} \}$ we used 50 - 200 operators. We performed calculations for longitudinal momenta $p_{\parallel} = 2 \pi q / l$, with $q=0,1,2$ and we span over $l \sqrt{\sigma}=1.4 - 7$. All our $SU(N>3)$ calculations are focused only on flux-tubes that carry a single unit of chromo-electric flux ($N$-ality $k=1$). Findings on the flux-tube spectrum with values of $N$-ality $k=2$ will be a matter of investigation in a future article.

The structure of these proceedings is the following. First, we provide a brief description of the lattice setup, by explaining the quantum numbers relevant for the extraction of the flux-tube spectrum as well as their encoding onto the operators used to materialize the calculation. Then we present the effective string theoretical descriptions suitable for approximating the spectrum of the flux-tube. Subsequently, we present our lattice results and then we conclude.

\vspace{-0.25cm}
\section{Lattice calculation}
\label{sec:lattice_calculation}
\vspace{-0.25cm}
We define the $SU(N)$ gauge theory on a $D=4$ Euclidean space-time lattice which has been compactified along all directions with volume
$L_{\parallel} \times L_{\perp_1} \times L_{\perp_2} \times L_{T}$. The length of the flux-tube is equal to $L_{\parallel}$, while $L_{\perp_1}$, 
$L_{\perp_2}$ and $L_{T}$ were chosen to be large enough to avoid finite volume effects. We choose $L_{\perp_1} = L_{\perp_2}  = L_{\perp}$ uniformly so that we ensure rotational symmetry around the flux axis. We perform Monte-Carlo simulations using the standard Wilson plaquette action $S=\sum_{\square} \beta \left[ 1-\frac{1}{N}{\rm Re}{\rm Tr}(U_{\square}) \right]\,,$ with inverse coupling $\beta=\frac{2N}{g^2(a)}$. In order to keep the value of the lattice spacing $a$ approximately  fixed for different values of $N$ we keep
the 't Hooft coupling $\lambda(a)=N g^2(a)$ approximately fixed, so that $\beta \propto N^2$. The simulation algorithm we use combines standard heat-bath and over-relaxation steps in the ratio 1:4; these are implemented by updating $SU(2)$ subgroups using the Cabibbo-Marinari algorithm. To measure the spectrum of energies we use the variational technique also called the Generalized Eigen-Value Problem (GEVP)
(e.g.~see \cite{variational} and its references).

The energy states of the closed flux-tube in $D=3+1$ are characterised by the irreducible representations of the two-dimensional lattice rotation symmetry around the principal axis denoted by $C_4$~\cite{Kuti1}. The above group is a subgroup of $O(2)$ corresponding to rotations by integer multiples of $\pi/2$ around the flux-tube propagation axis. As a result, values of angular momenta that differ by an integer multiple of four are indistinguishable on the lattice. This splits the Hilbert space in four orthogonal sectors, namely: $J_{{\rm mod}\, 4}=0$, $J_{{\rm mod}\, 4}=\pm 1$,  $J_{{\rm mod}\, 4}=2$. Furthermore, parity $P_\perp$ which is associated with reflections around the axis $\hat{\perp}_1$ can be used to characterise the states. Applying $P_\perp$ transformations, flips the sign of $J$. Therefore, one can choose a basis in which states are characterised by their value of $J$ ($J= \pm$), or by their value of $|J|$ and  $P_\perp$. We adopt the latter. In the continuum, states with $J \neq 0$ are parity degenerate, however, on the lattice this holds only for the odd values of $J$. In practice, we describe our states with the following $5$ irreducible representations $A_{1}$, $A_{2}$, $E$, $B_{1}$ and $B_{2}$ of $C_4$ group whose $J$ and $P_\perp$ assignments are: $\left\{ A_1: \, |J_{{\rm mod} \ 4}|=0, \ P_\perp=+\right\}, \left\{A_2:  \, |J_{{\rm mod} \ 4}|=0, \  P_\perp=-\right\}$, $\left\{ E: |J_{{\rm mod} \ 4}|=1, \  P_\perp=\pm\right\}$,  $\left\{ B_1: \, |J_{{\rm mod} \ 4}|=2, \ P_\perp=+\right\}$ and $\left\{B_2:  \, |J_{{\rm mod} \ 4}|=2, \ P_\perp=-\right\}$.

Furthermore, there are two additional quantum numbers which can be proven useful for the description of the string states. These are the longitudinal momentum $p_{||}$ carried by the flux-tube along its axis (which is quantized in the form $p_{||}=2\pi q/L_{||}; q\in Z$) and the parity $P_{||}$ with respect to reflections across the string midpoint. Since $P_{||}$ and $p_{||}$ do not commute, we can use both to  simultaneously characterise a state only when $q=0$. The energy does not depend on the sign of $q$, hence, we only focused on those with $q\ge 0$.

\begin{figure}
    \centering
    \vspace{-0.5cm}
    \includegraphics[height=7cm]{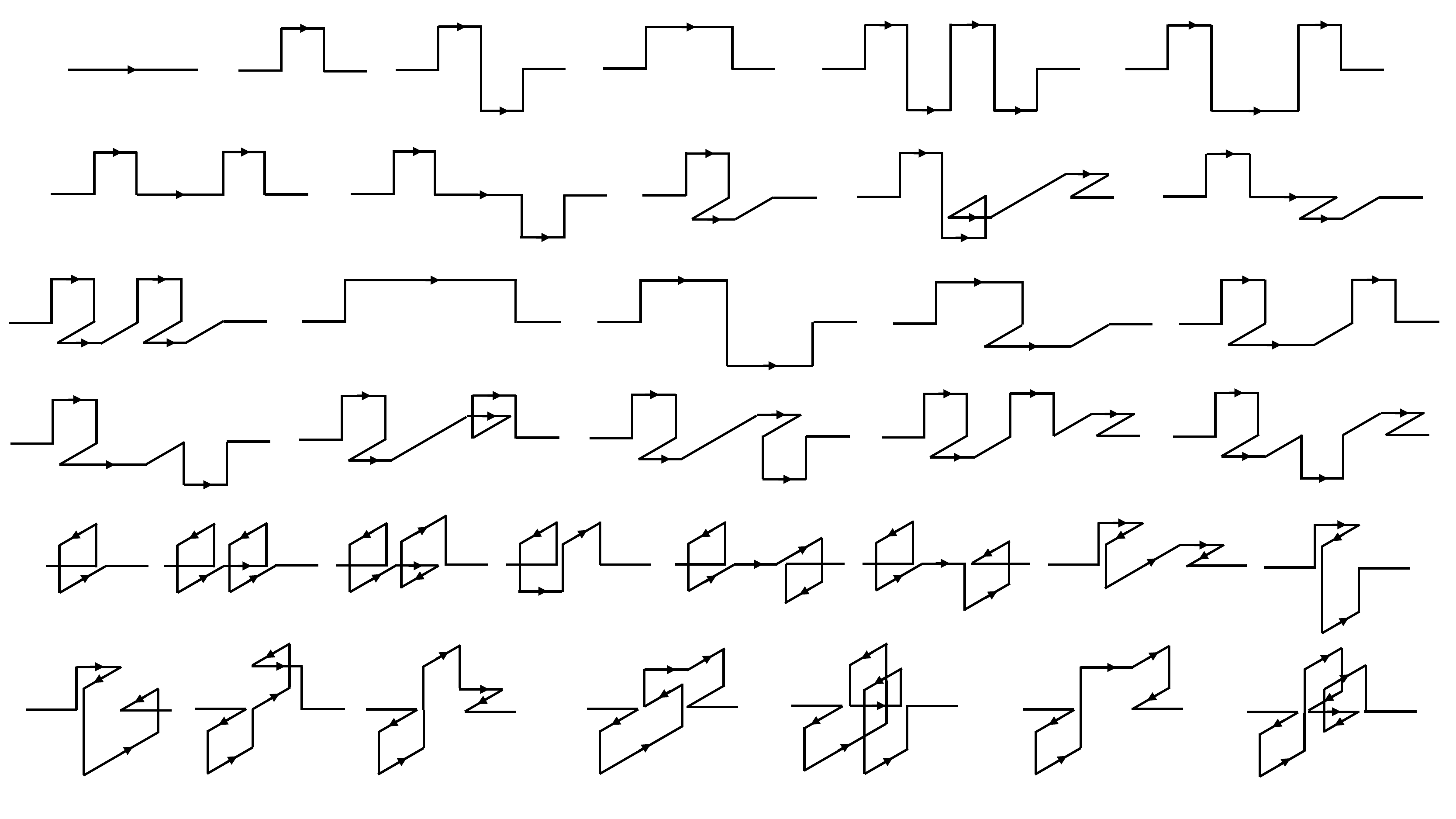}
    \caption{All the different paths used for the construction of the torelon operators.}
    \label{fig:paths}
\end{figure}

Flux-tube energies are extracted by making use of correlation matrices $C_{ij} = \langle \phi_i^{\dagger} (t) \phi_j (0) \rangle$ with $i,j=1...N_{\rm op}$ in combination with GEVP where $N_{\rm op}$ the number of operators. We construct operators $\phi_i$ which encode shapes that lead to particular values of $J,P_\perp,P_{||},$ and $q$. We do so by choosing 
linear combinations of Polyakov loops the paths of which consist of  various  transverse deformations and various smearing and blocking levels~\cite{variational}. All the transverse paths used for the construction of the operators are shown in Figure~\ref{fig:paths} and all together, including smearing and blocking levels, form a basis of around $N_{\rm op} =1000$ operators with approximately $50 - 200$ for each different irreducible representation. To build an operator which encodes a certain value of angular momentum $J_{{\rm mod} \ 4}$ we begin the construction with a sub-operator $\phi_{\alpha}$ which has a deformation extending in angle $\alpha$ within the plane of transverse directions. Then we repeat the same procedure by rotating the sub-operator by integer values of $\pi/2$. Finally, we can construct the operator $\phi(J)$ belonging to a specific representation of $C_{4}$ by using the formula $\phi(J)= \sum_{n=1,2,3,4} e^{iJ n \frac{\pi}{2}} \phi_{n \frac{\pi}{2}}\,.$ Thus $\phi(0)$ belongs to $A_1$ and $A_2$, $\phi(1)$ to $E$ and, finally, $\phi(2)$ to $B_1$ as well as $B_2$. Lastly, it is required to encode certain values of $P_{\perp}$ and $P_{||}$ by summing and subtracting reflections of the initial sub-operator $\phi(J)$ over the transverse and parallel parity planes. Such an example is pictorialized in Equation~\ref{eq:example} for an operator with $J_{\rm mod \ 4}=0$.
\begin{eqnarray}
 \phi= {\rm Tr}\left[ \parbox{12.5cm}{\rotatebox{0}{\includegraphics[width=12.5cm]{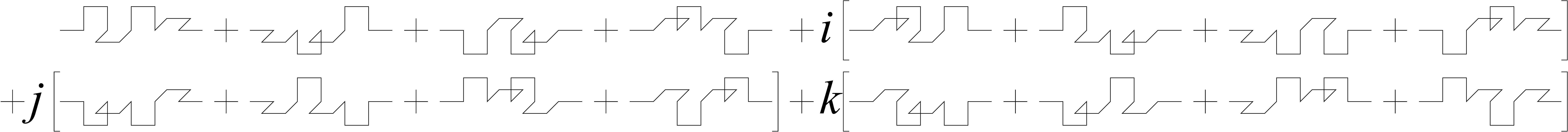}}} \ \right]\,.
\label{eq:example}
\end{eqnarray}
For the combination $i=j=k=+1$, $\phi$ projects onto $\{ A_1, P_{||}=+ \}$, for $i=+1,j=k=-1$ onto $\{ A_2, P_{||}=+ \}$, for $i=-1,j=+1,k=-1$ onto $\{ A_1, P_{||}=- \}$ and finally, for $i=j=-1,k=+1$, onto $\{ A_2, P_{||}=- \}$.

\section{The effective string theory}
\label{sec:effective_string}
\vspace{-0.25cm}
Imagine a flux-tube as a string with length $l=aL_{||}$ winding around the spatial torus. Imposing fixed spatial position for the string  spontaneously  breaks  translation  symmetry. Hence, we expect $D-2$ Nambu-Goldstone massless bosons to appear at low energies. Such bosons reflect the transverse fluctuations of the flux-tube around its classical configuration. We would thus, expect a low energy Effective String Theory (EST) describing the flux-tube spectrum at large enough lengths. Of course a flux-tube is not an infinitesimally thin string, it lives in $SU(N)$ manifold and presumably has an intrinsic width $l_w\propto 1/\sqrt{\sigma}$. We would therefore expect that the spectrum of the flux-tube consist not only of string like states but also of massive excitations. Below, we describe the current theoretical predictions for the excitation spectrum of the Nambu-Goldston bosons as well as an approach to explain the existence of massive resonances on the world-sheet of the flux-tube.
\subsection{The Goddard–Goldstone–Rebbi–Thorn string}
\label{sec:GGRT}
At this subsection we describe the spectrum of the Goddard-Goldstone-Rebbi-Thorn (GGRT)~\cite{Goddard} or in simpler words the Nambu-Goto (NG)~\cite{NGpapers} closed string. NG string describes non-critical relativistic bosonic strings. One can extract the GGRT spectrum by performing light-cone quantization of the closed-string using the NG action or equivalently the Polyakov action. NG action is the area of the world-sheet swept by the propagation of the string along the time direction. This model is Lorentz invariance only in $D=26$ dimensions. Nevertheless, for reasons that we now understand better~\cite{Dubovsky:2013gi} NG can also describe adequately the spectrum of strings in $3 \ {\rm and} \ 4$ dimensions. The spectrum of the GGRT string is given by the expression:
\begin{equation}
{E_{N_L,N_R}(q,l)}
= \sigma l \sqrt{
1 
+
\frac{8\pi}{(l \sqrt{\sigma})^2} \left(\frac{N_L+N_R}{2}-\frac{D-2}{24}\right)
+
\left(\frac{2\pi q}{(l\sqrt{\sigma})^2}\right)^2}\,,
\label{eqn_EnNG}
\end{equation}
where $2\pi N_{L(R)}/l$ the total energy and momentum of the left(right) moving phonons with $N_L = \sum_k \sum_{n_L(k)} k(n^+_L(k)+ n^-_L(k))$ and
 $N_R = \sum_k \sum_{n_R(k)} k(n^+_R(k)+ n^-_R(k))$. $n^{\pm}_{L(R)}(k)$ is the number 
of left(right) moving phonons of momentum $p_k = 2\pi k/l$, $k=0,1,2,\dots$ and angular momentum  $\pm 1$. If $p_{||}=2\pi q/l$ is the total longitudinal momentum of the string then, since the phonons provide that momentum, we must have $N_L - N_R = q$. The angular momentum around the string is given by $J = \sum_{k,n_L(k),n_R(k)} n^+_L(k) + n^+_R(k) - n^-_L(k) - n^-_R(k)$.


\subsection{Lorentz invariant string approaches}
\label{sec:Lorentz}
 Systematic ways to study Lorentz invariant EST that would describe the QCD flux-tube were pioneered by  L\"uscher, Symanzik, and Weisz in \cite{Luscher} (static gauge) as well as by Polchinski and Strominger in~\cite{Pol} (conformal gauge). Such EFT approaches produce predictions for the energy of states as an expansion in $1/l\sqrt{\sigma}$. Terms in this expansion that are of $O(1/l^p)$ are generated by $(p+1)$-derivative terms in the EST action whose coefficients are a priori arbitrary Low Energy Coefficients (LECs). Interestingly, these LECs were shown to obey strong constraints that reflect a non-linear realization of Lorentz symmetry \cite{LW,Meyer,AK}, and so to give parameter free predictions for certain terms in the $1/l$ expansion.
 
 The EST approaches can be characterised by the way one performs the gauge fixing of the embedding coordinates on the world-sheet. This can be either the static gauge~\cite{Luscher,LW,AK} or the conformal gauge~\cite{Pol,Drummond:2004yp,HariDass:2009ub} with both routes leading to the same results. The starting point of building the EST is the leading area term which gives rise to the linearly rising potential for large strings. Subsequently comes the Gaussian action which is responsible for the $\propto 1/l$ L\"uscher term with universal coefficient depending only on the dimension $D$. As a next step one adds the 4-derivative terms which yield a correction on the energy spectrum proportional to $1/l^3$ with a universal coefficient that also depends on the dimension $D$. One can include the $6-$derivative terms and show that for $D=3$ they yield the fourth universal term proportional to $1/l^5$ in the energy spectrum, while for general states in $D=4$, the coefficient of the $O(1/l^5)$ term is not universal. Nonetheless, the energy just for the ground state in the $D=4$ case is universal. Summarizing the above information, the spectrum is given by 
 \begin{eqnarray}
  E_n(l) =  \sigma l + \frac{4 \pi}{l}\bigg(n-\frac{D-2}{24}\bigg)  -  \frac{8 \pi^2}{\sigma l^3}\bigg(n-\frac{D-2}{24}\bigg)^2  +   \frac{32 \pi^3}{\sigma^2 l^5}\bigg(n-\frac{D-2}{24}\bigg)^3   + {O}(l^{-7}).
\label{eq:AharonyKarzbrun}
\end{eqnarray}
Since we think of the GGRT model as an EST, which may be justified only for long strings \cite{Olesen}, one can expand the associated energy for $l\surd \sigma\gg 1$. The result of the expansion is the same as Equation~\ref{eq:AharonyKarzbrun}. For simplicity we set $q=0$, and  $n = (N_L + N_R)/2$. 

\subsection{The topological term action}
\label{sec:axion}
In 2013, Dubovsky {\it et al.} worked out an approach for extracting the spectrum of the flux-tube for short as well as for long lengths. The idea was based on the fact that the GGRT string provides the best approximation for the flux-tube spectrum and that Equation~\ref{eqn_EnNG} can be re-expressed as $E_{N_L,N_R} = \sqrt{\sigma} {\cal E} (p_k/\sqrt{\sigma},1/l \sqrt{\sigma})$ where $p_k$ are the momenta of individual phonons in units of $2 \pi / l$ comprising the state quantised. The naive expansion in $1/ l \sqrt{\sigma}$ is the combination of two different expansions; the first is an expansion in the softness of individual quanta compared to the string scale, i.e. in $p_k / \sqrt{\sigma}$ and the second expansion is a large volume expansion, i.e. an expansion in $1 / l \sqrt{\sigma}$. To disentangle the two expansions the following procedure is being followed. First, one calculates the infinite volume $S$-matrix of the phonon collisions. This is done perturbativelly given that the center of mass energy of the colliding phonons is small in string units; this is called the momentum expansion. Followingly, the authors extract the finite volume energies from this $S$-matrix by using approximate integrability and the Thermodynamic Bethe Ansatz (TBA). This allows to extract the winding effects on the energy from virtual quanta traveling around the circle as well as the winding corrections due to phonon interactions.

The authors demonstrated that when a state has only left-moving phonons the GGRT winding corrections in the energy spectrum are small and, thus, one expects the spectrum to be close to that of the free theory. On the contrary, for states containing both left- and right-movers, energy contributions are larger. The above picture is in a good agreement with most of the states in $D=4$ but fails to explain the anomalous behaviour of the pseudoscalar level firstly reported in \cite{Athenodorou:2010cs} suggesting that an additional ingredient is required in order to describe such excitations. The most straightforward way to do this is the introduction of a massive pseudoscalar particle $\phi$ on the world-sheet. The leading interaction compatible with non-linearly realized Lorentz
invariance for such a state is a coupling to the topological invariant known as the self-intersection number of the string $S_{\rm int} = \frac{\alpha}{8 \pi} \int d^2 \sigma \phi K^{i}_{\alpha \gamma} K^{j \gamma}_{\beta} \epsilon^{\alpha \beta} \epsilon_{ij} \,$
with $K^{i}_{\alpha \gamma}$ being the extrinsic curvature of the world-sheet, $\alpha$ the associated coupling and $\sigma^{i}$, $i=1,2$ the world-sheet coordinates. Adapting the above interaction term to our old results for $SU(3)$, $\beta=6.0625$ yields a mass of $m_{\phi}/\sqrt{\sigma} \simeq 1.85^{+0.02}_{-0.03}$ and a coupling of $\alpha = 9.6  \pm 0.1$.


\section{Results}
\label{sec:results}
\vspace{-0.25cm}
In this work we present results for the closed flux-tube spectrum extracted from calculations on five different gauge groups. The above consist of 
$N=3$ at $\beta=6.0625$ ($a\simeq 0.09 {\rm fm}$) and $\beta=6.338$ ($a\simeq 0.06 {\rm fm}$), of $N=5$ at $\beta=17.630$ ($a\simeq 0.09 {\rm fm}$) and $\beta=18.375$ ($a\simeq 0.06 {\rm fm}$) as well as for $N=6$ at  $\beta=25.550$ ($a\simeq 0.09 {\rm fm}$). Critical slowing down~\cite{Athenodorou:2021qvs,Athenodorou:2020ani}, as one moves towards the continuum and the large-$N$ limit, prohibits the investigation of gauge groups with $N \geq 6$ and $a < 0.09 {\rm fm}$. Nevertheless, the above configuration of measurements is enough to determine whether significant lattice artifacts as well as $1/N^2$ corrections are affecting our statistically more accurate $N=3$ calculations. As a matter of fact our investigation demonstrates that such effects are of minor importance and do not play a significant role in the interpretation of the spectrum. The energy spectrum we extracted is compared to the predictions of the GGRT string. Namely, we fit the absolute ground state ($|J_{\rm mod \ 4}|^{P_{\perp} P_{||}} =0^{++}$) for all calculations using Equation~\ref{eqn_EnNG} as a function of the length for $l\sqrt{\sigma} > 2.5$ and extract the string tension $a \sqrt{\sigma}$. Once the string tension is known then Equation \ref{eqn_EnNG} can be used as a parameter free prediction for higher string excitations with $N_L + N_R > 0$. 

\subsection{The energy spectrum for $q=0$ and the world-sheet axion}
\label{sec:resultsq0}
\begin{figure}
    \centering
    \vspace{-1.5cm}
    \hspace{-3.5cm}
    \includegraphics[height=8.1cm]{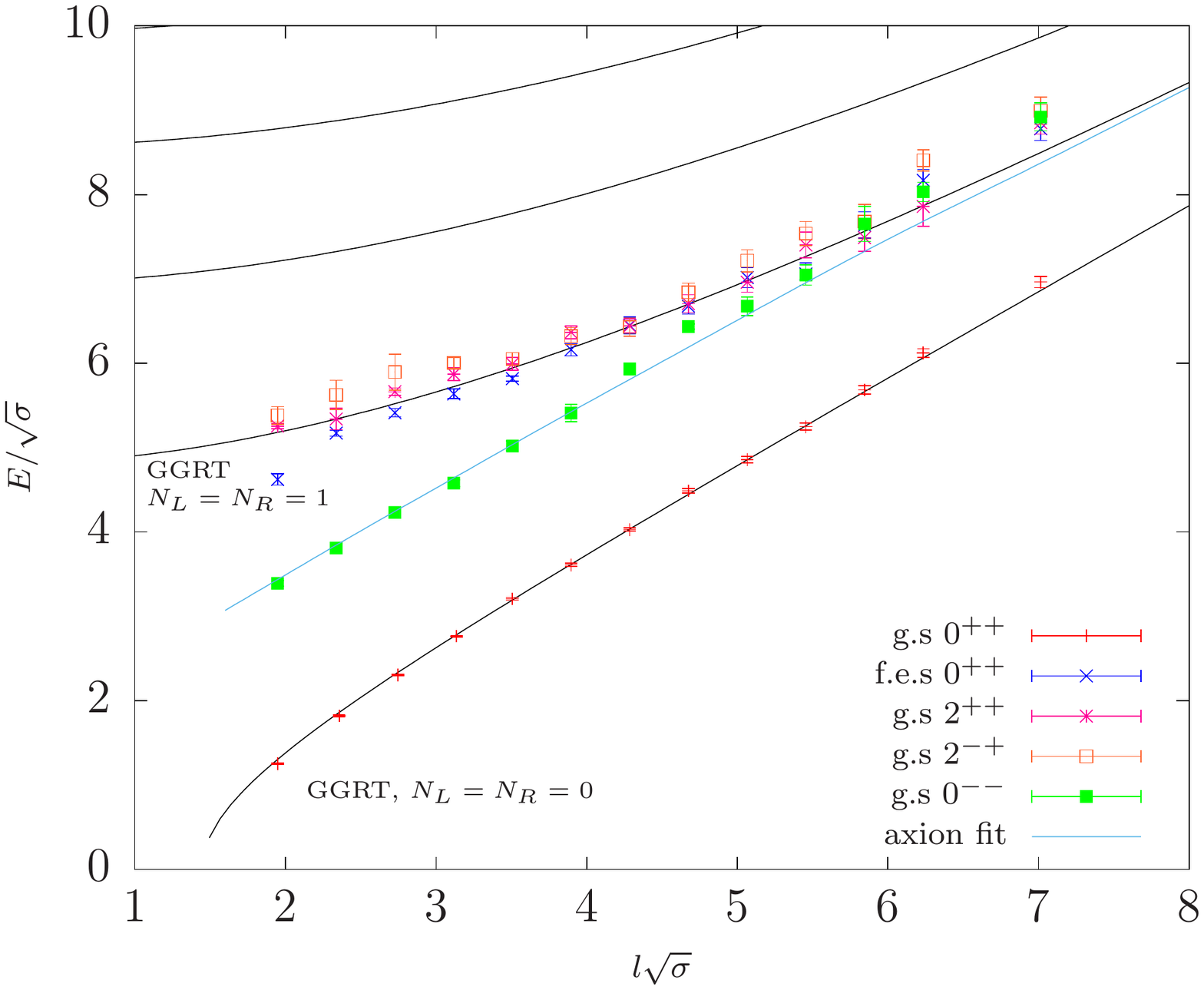} \hspace{-3cm} \includegraphics[height=8.1cm]{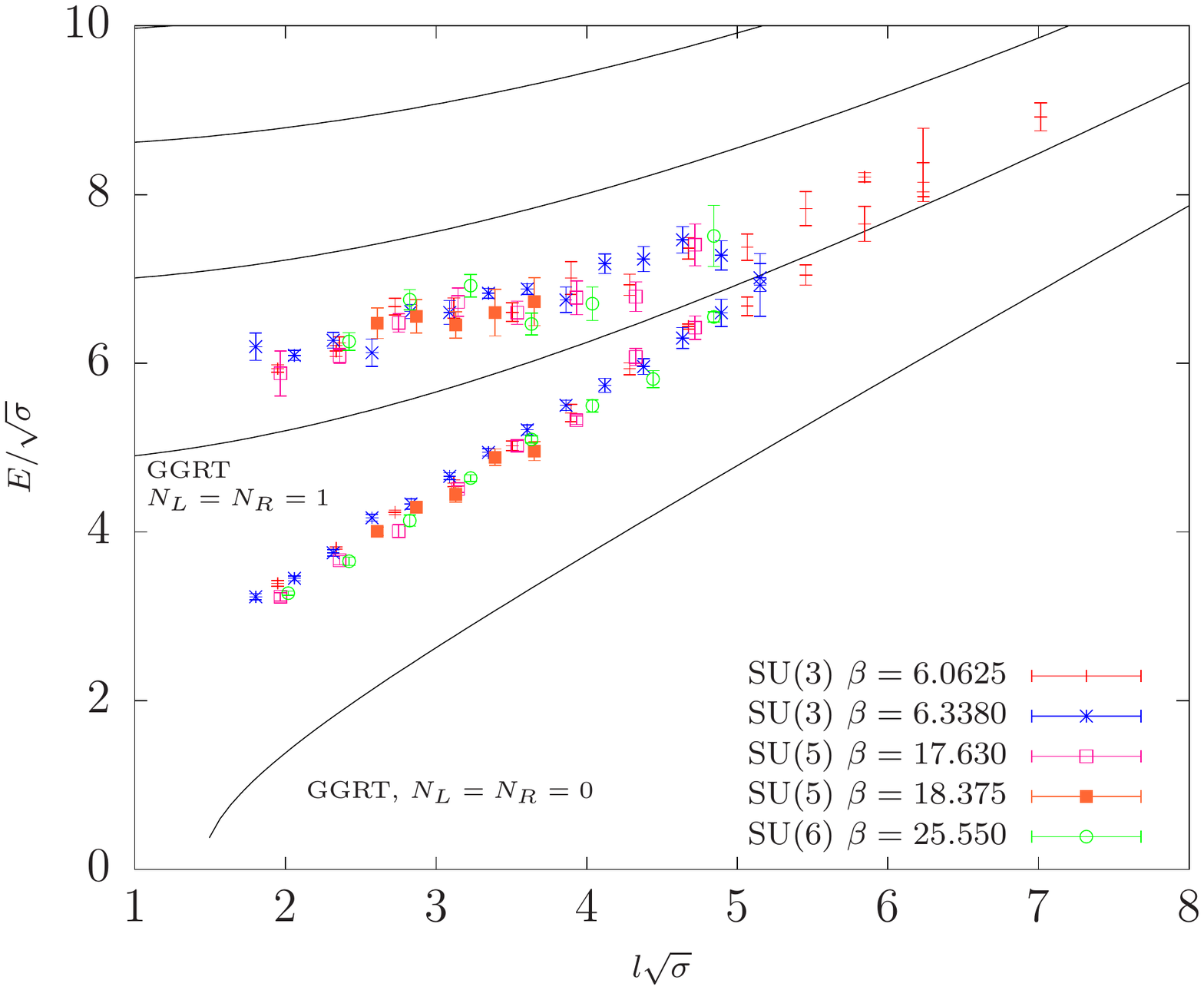} \hspace{-4.5cm}
    \vspace{-1cm}
    \caption{ \underline{Left Panel:} The spectrum of the absolute ground state and first excited state for $|J_{\rm mod \ 4}|^{P_{\perp}, P_{||}}=0^{++}$ as well as the ground states for $2^{++}$, $2^{-+}$ and the "anomalous"  $0^{--}$ for $q=0$ and $SU(3)$ at $\beta=6.0625$; the black lines correspond to the GGRT predictions and the light blue line to the prediction of the EFT with the axionic part of the action included within. \underline{Right Panel:}. The energies of the ground state and first excited state for $|J_{\rm mod \ 4}|^{P_{\perp}, P_{||}}=0^{--}$, $q=0$ for all gauge groups considered in this work.}
    \label{fig:first}
\end{figure}

\begin{figure}
    \centering
    \vspace{-1.5cm}
    \includegraphics[height=8.1cm]{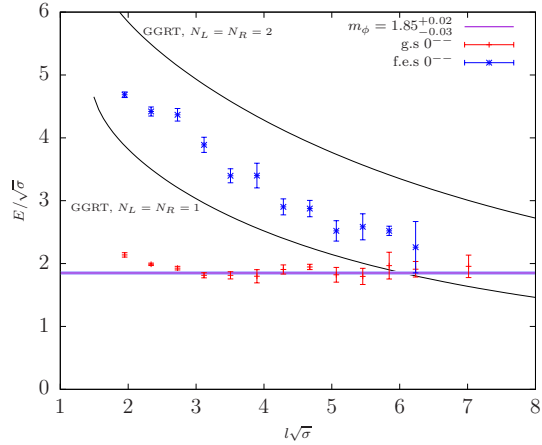}
    \vspace{-1cm}
    \caption{The energy levels of the ground and first excited states for a closed flux-tube with quantum numbers $0^{--}$, $q=0$ and the zeroth energies subtracted for $SU(3)$, $\beta=6.0625$. The horizontal purple band corresponds to the mass of the axion as this has been extracted in \cite{Dubovsky:2013gi}.}
    \label{fig:second}
\end{figure}

We begin by presenting our results for the $q=0$ longitudinal momentum sector in Figures~\ref{fig:first}, \ref{fig:second} and \ref{fig:third}. In the left panel of Figure~\ref{fig:first}, the lowest energy level corresponds to the absolute ground state $|J_{\rm mod \ 4}|^{P_{\perp} P_{||}} =0^{++}$ which is used to set the scale of the NG string, hence, the nearly perfect agreement with the GGRT string. Furthermore in the left panel of Figure~\ref{fig:first}, we plot the first excited state of $0^{++}$ as well as the ground states of $2^{++}$, $2^{-+}$ and $0^{--}$ for $SU(3)$ at $\beta=6.0625$. We compare the above data with the GGRT prediction for $N_R=N_L=1$. This string state is expected to be four-fold degenerate with levels with continuum QNs $0^{++}$, $0^{--}$, $2^{++}$ and $2^{-+}$ . While $0^{++}$, $2^{++}$ and $2^{-+}$ flux-tube excitations appear to exhibit small deviations for short values of $l\sqrt{\sigma}$ and for larger strings become consistent with GGRT, $0^{--}$ ground state appears to demonstrate significant deviations from the GGRT string. In the right panel of Figure~\ref{fig:first} we present the ground and in addition the first excited state with QNs $0^{--}$ for all gauge groups considered in this work. It appears that both states are only mildly affected by lattice artifacts and $1/N^2$ corrections. The $0^{--}$ ground state appears to have characteristics of a resonance i.e. a constant mass term coupled to the absolute ground state. This is more obvious by subtracting the absolute ground state $0^{++}$ where this excitation exhibits a plateau; this  is presented in Figure~\ref{fig:second} for $SU(3)$ at $\beta=6.0625$. As has already being explained in Section~\ref{sec:axion} this state can be well interpreted as an axion on the world-sheet of the flux-tube with an associated mass of $m/\sqrt{\sigma} = 1.85^{+0.02}_{-0.03}$ for $SU(3)$ at $\beta=6.0625$; This value is in good agreement with the plateau in Figure~\ref{fig:second}. If the $0^{--}$ flux-tube ground state corresponds to the axion, the next excitation level would correspond to the string state with $N_L=N_R=1$ rather than  $N_L=N_R=2$. As one can see in the right panel of Figure~\ref{fig:first} and Figure~\ref{fig:second} the $0^{--}$ first excitation state does not approach the GGRT $N_L=N_R=2$ state but instead it slowly approaches the $N_L=N_R=1$ string state. This strengthens the scenario of $0^{--}$ ground state being 
the world-sheet axion.

\subsection{A bound state of two axions?}
\label{sec:two_axions}
\begin{figure}
    \centering
    \vspace{-1.5cm}
    \hspace{-3cm}
    \includegraphics[height=8cm]{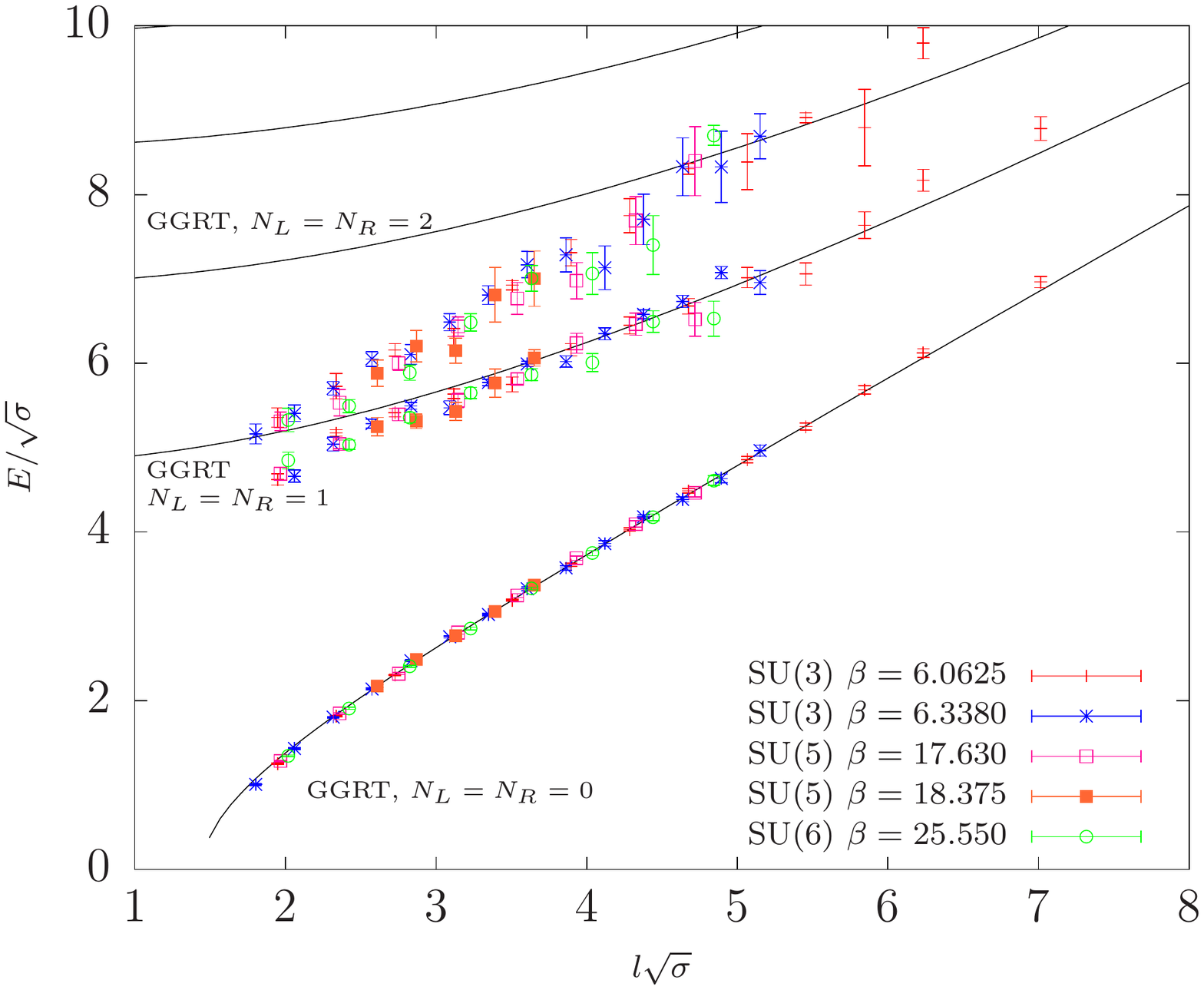} \hspace{-3cm}
    \includegraphics[height=8cm]{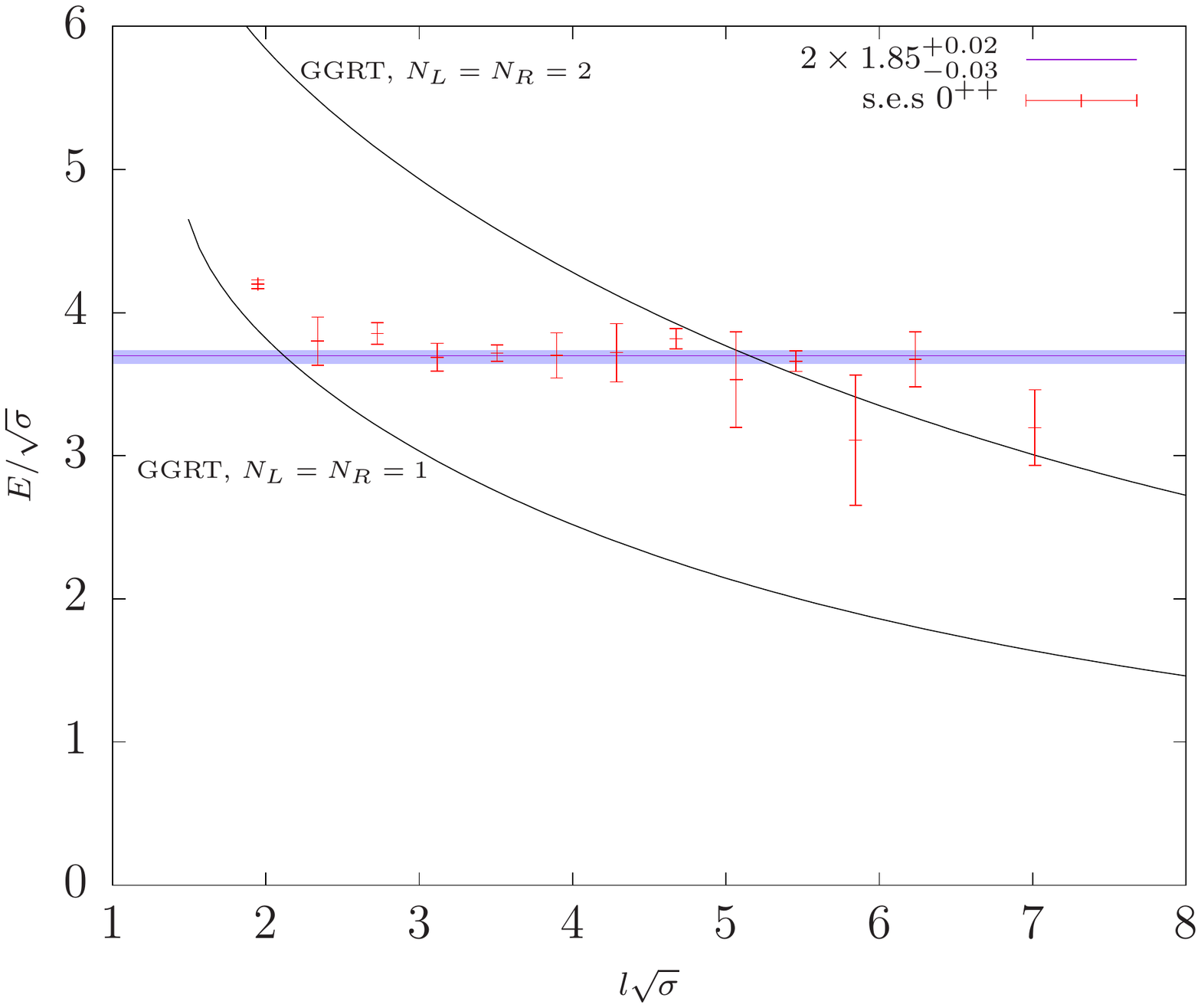} \hspace{-4cm}
    \vspace{-1cm}
    \caption{{\underline{Left Panel}}: The ground, first excited and second excited state for $|J_{\rm mod \ 4}|^{P_{\perp}, P_{||}}=0^{++}$ and all the gauge groups used in this work. {\underline{Right Panel}}: The second excited state for $|J_{\rm mod \ 4}|^{P_{\perp}, P_{||}}=0^{++}$ with the absolute ground state being subtracted for $SU(3)$ at $\beta = 6.0625$.}
    \label{fig:third}
\end{figure}

In the left panel of Figure~\ref{fig:third} we present the second excitation state with QNs $ 0^{++}$. It is well mentioned that above this energy level we get a plethora of states which reflect the multifold degeneracy of the GGRT string for $N_L = N_R = 2$. Strikingly, this state appears to exhibit the same resonance behaviour as the $0^{--}$ ground state i.e. it appears as a constant term coupled to the absolute ground state. This is more obvious if we subtract from this energy level the contribution of the absolute ground state as this appears in the right panel of Figure~\ref{fig:third}. Namely, we observe that this is in agreement with a resonance of mass twice that of the axion. This raises the question whether such a relation is accidental or it has some deeper interpretation. A reasonable expectation would be that this state is a bound state of two axions with a very low binding energy; this scenario is in agreement with the quantum numbers of the state.

\subsection{The $q \ne 0$ sector and the world-sheet axion}
\label{sec:nonzeromomentum}

\begin{figure}
    \centering
    \vspace{-1.5cm}
    \hspace{-3cm}
    \includegraphics[height=8cm]{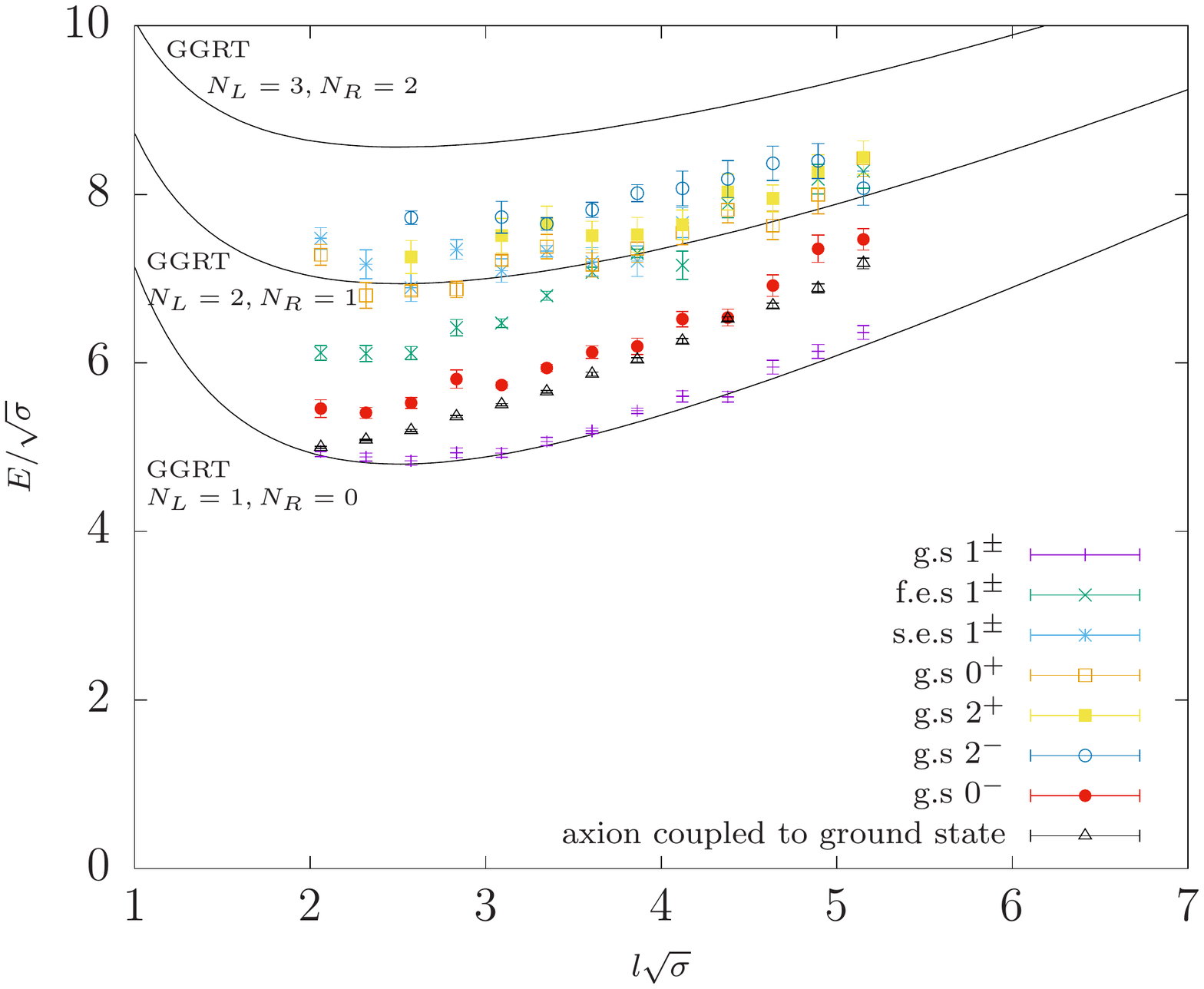} \hspace{-3cm}
    \includegraphics[height=8cm]{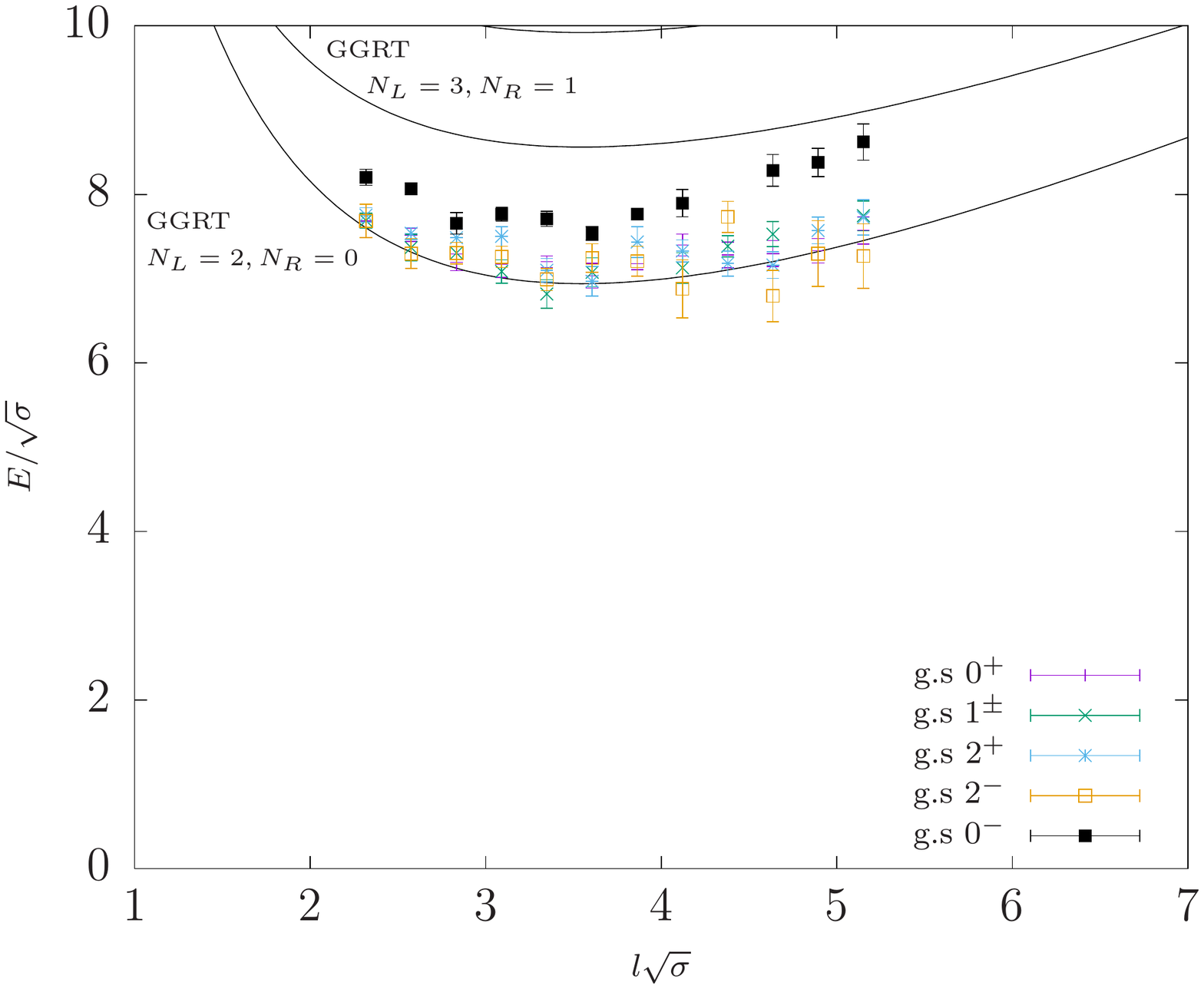} \hspace{-4cm}
    \vspace{-1cm}
    \caption{{\underline{Left Panel}}: The ground, first excited and second excited $1^{\pm}$ states as well as the ground states $0^+$, $0^-$, $2^+$, $2^-$ for a flux-tube with $q=1$ in $SU(3)$, $\beta=6.338$. {\underline{Right Panel}}: The ground states with QNs $0^{+}$, $0^-$, $1^{\pm}$, $2^+$, $2^-$ for a flux-tube with $q=2$ at $SU(3)$, $\beta = 6.338$.}
    \label{fig:fourth}
\end{figure}

In this section we present our results for the $q=1$ and $q=2$ momentum sectors. In the left panel of figure~\ref{fig:fourth} we demonstrate the spectrum  for $q=1$, $SU(3)$ and $\beta=6.338$. Since, the string ground state $N_L=1$, $N_R=0$ can only be created by a single phonon, it has $J=1$. The flux-tube ground state with quantum numbers $1^{\pm }$, $q=1$ appears to be in good agreement with the prediction of the GGRT string. This is in accordance with the results of Ref~\cite{Dubovsky:2013gi}. The next string excitation level, corresponding to $N_L=2$ and $N_R=1$ should be seven-fold degenerate. This should consist of one $0^{+}$, one $0^{-}$, three $1^{\pm}$, one $2^{+}$ and a $2^-$ state. In the left panel of Figure~\ref{fig:fourth} we show the flux-tube ground state with QN $2^+$, the ground state with $2^-$, the ground state for $0^+$ as well as the first and second excited states with $1^{\pm }$. All the above five states appear to cluster around the GGRT prediction. Furthermore, we demonstrate the ground state for $0^{-}$ which appears to exhibit large deviations from the GGRT string. Since, this state has the same QNs as the pseudoscalar massive excitation the first assumption one could make is that this reflects to the axion. A naive comparison of this state with a relativistic sum of the absolute ground state plus an axion with momentum $2 \pi / l$ is provided in the same figure, demonstrating an approximate agreement with our data for large flux-tubes. This strengthens the scenario of this state being the axion.  

In the right panel of figure~\ref{fig:fourth} we show results for $q=2$, $SU(3)$ and $\beta=6.338$. The string ground state $N_L=2$, $N_R=0$ is expected to be four-fold degenerate. Namely, it is expected to be occupied by states with QNs $0^{+}$, $1^{\pm}$, $2^{+}$ and $2^{-}$. We, thus, extract the flux-tube ground states with the above QNs and observe that they all cluster around the GGRT prediction. The next string excitation level is multi-fold degenerate and should also include a $0^{-}$ state which encodes the QN of the axion. We extract the flux-tube ground state with QNs $0^-$ and we observe a very similar behaviour as for the case of $q=1$; namely it diverges greatly from the GGRT prediction. 

\vspace{-0.25cm}
\section{Conclusions}
\label{sec:conclusions}
\vspace{-0.25cm}
We have improved extensively the extraction of and, thus, our knowledge on the spectrum of the closed flux-tube (torelon). Clearly, the majority of the states appearing in the spectrum have a string-like character, in the sense they can be adequately approximated by a low energy effective string theory. In addition a small sector of the excitation spectrum appears to be massive resonances which can be interpreted as an axion on the bulk of the theory. This is justified by the resonance character of the $0^{--}$, $q=0$ ground state which appears to be an axion coupled to the string's absolute ground state, by the $0^{++}$ second excited state which can be interpreted as a bound state of two axions with a very low binding energy coupled to the absolute ground state as well as by the $0^-$ $q=1,2$ ground states which also have an axion character. Finally, and not presented in this manuscript, states with axionic character can also be identified in other irreducible representations such as $|J_{\rm mod \ 4}|^{P_{\perp}} = 1^{\pm}$; this will be a matter of discussion in our longer write up~\cite{AthenodorouTeperNew}.

\vspace{-0.25cm}
\section*{Acknowledgements}
\vspace{-0.25cm}
We would like to thank S. Dubovsky, D. Giataganas, V. Gorbenko, J. Sonnenschein, E. Kiritsis and K. Hashimoto for interesting discussions. As this work was being completed, AA participated in the {\it Large-$N$ theories and strings: conformal, confining, and holographic} Workshop at the Princeton Center for Theoretical Physics, where there were many talks relevant to confining flux-tubes. AA is indebted to the participants for useful discussions. AA has been financially supported by the European Union's Horizon 2020 research and innovation programme ``Tips in SCQFT'' under the Marie Sk\l odowska-Curie grant agreement No. 791122. Numerical simulations have been carried out in the Oxford Theoretical Physics cluster.

\end{document}